\documentclass[1p,longtitle]{elsarticle}

\usepackage{lineno,hyperref}
\usepackage{amsmath}
\journal{Physica A: Statistical Mechanics and its Applications}

\begin{document}

\begin{frontmatter}

\date{}

\title{A Markovian  approach to the Prandtl-Tomlinson  frictional model}


\author[1]{D. Lucente}
\author[2]{A. Petri}
\author[3]{A. Vulpiani}

\address[1]{Univ Lyon, ENS de Lyon, Univ Claude Bernard de Lyon, CNRS, Laboratoire de
	Physique,F-69342 Lyon, France}
\address[2]{CNR - Istituto dei sistemi Complessi,  Unit\`a Sapienza, P.le A. Moro 2, 00185 Roma, Italy}
\address[3]{Dipartimento di Fisica, Universit\`a Sapienza, P.le A. Moro 2, 00185 Roma, Italy}

\begin{abstract}
We consider the Prandtl-Tomlinson model in the case of a constant driving force and in the presence of thermal fluctuations. We show that the system 
dynamics is well reproduced by
a simplified description obtained through a Markov process, even in the case of potentials with several minima. After estimating the chain parameters by numerical simulation, we compute  the average velocity and friction at varying driving  force and temperature. Then we  take advantage  of this approach  for calculating the  entropy produced by the system and,  in the case of a single minimum potential, to derive its explicit relation with the external force and the mobility at low temperatures.  We observe that the coefficient relating  the entropy production to the force  is not a monotonic function of the temperature.
\end{abstract}


\end{frontmatter}


\section{Introduction}

Among the simplified models for solid-on-solid friction, the now so called Prandtl-Tomlinson (PT) model  has recently seen renewed interest because of its suitability to the description of atomic force microscopy \cite{Dong2011,Vanossi2013}.  As already pointed out in \cite{Popov2012} this model mainly originates from a work of
L.  Prandtl  \cite{Prandtl1928} and has very limited  relevance with the model for  friction devised  by G. Tomlinson  \cite{Tomlinson1929} (a
summary of its history and different contributions can be found in  several reviews, e.g. \cite{Popov2012} and \cite{spikes2015}).  In its present acceptation the PT model 
describes the dissipative motion of  a particle on a periodic substrate, and in its  more general formulation it  is defined by  the Langevin equation
\begin{equation}
m\ddot{x}=f(x)-\partial_x U(x)-\gamma\dot{x} +\sqrt{2 \gamma k_B T} \xi(t),
\label{lang}
\end{equation}
being   $m$ is the mass of the probe particle, $f(x)$ an external force acting on the probe,  and $\gamma$ a viscous coefficient;
$U$ is a periodic function, $U(x+L)=U(x)$, representing the substrate characterized from some crystalline order. Here   the effect of temperature, $T$, is given  by the last term which  derives its form from the Einstein's dissipation-fluctuation relation, linking the noise amplitude to the viscosity and the  Boltzmann constant $k_B$. 
Depending on the choice of $f(x)$, different  implementations  of the model are possible \cite{Popov2012}. While the case of a probe driven by a compliant harmonic force with constant drift,
$f(x)=-k(vt-x),$ has  been widely investigated   \cite{Fusco2005,Nakamura2005,Furlong2009,Dong2012,Wang2015}, here we focus on the case of a constant driving force, $f(x)=F$, which can be  relevant to the description of macroscopic interfaces and has been much less studied. Recent work on this case has focused mainly on the lowering of the critical energy barriers with increasing the driving force \cite{Furlong2015,Xu2019}.  

In this paper we describe  the dynamics of the PT model as a Markov chain where the relevant states are the potential minima.  By investigating  the stationary average speed and  friction  we  find that  such process can effectively well describe the system
 in a wide range of $F$ and $T$, allowing 
to  simplify the calculation of quantities of interest, such as the mobility and  the entropy production rate.  For illustration  we shall consider  both the case of a  substrate potential displaying a single minimum per period and that in which several periodic minima are present.

In Section 2 we briefly discuss the general reasons to adopt a description in terms of Markov chain, as well as the basic mathematical aspects and the numerical results. Section 3 is devoted to the investigation of friction and velocity, including an evaluation of the mobility,  whereas  Sec.  4  concerns  the entropy production.  Section 5 shortly investigates an instance  of a substrate displaying $m$ minima and reports the behavior of mobility and entropy production for the specific case of $m=4$.
Section 6 summarizes the main  results.

\section{Markov description: conceptual and numerical determination}

Often in experimental or numerical measurements, the state of a physical system cannot be determined with arbitrary accuracy. This implies the possibility (necessity) to introduce probabilistic methods using suitable stochastic description. In particular, a description in terms of Markov chains, which is quite natural and (relatively) simple, had been used with success in many different contexts \cite{bowman2013introduction}. Such an approach can be used just from the knowledge of experimental data and even in absence of a good model.

In  a Markov chain  \cite{Norris},  time  is discretized into intervals $\tau$ and the process is described by a numerable  set of  states $\lbrace s \rbrace $, with the assumption  that  the
conditional probability that the system at time $t_n$ is in the state $s_\alpha$, knowing the past, i.e.  the  states  at times $t_{n-1}$, $t_{n-2}$,  and so on,   depends only by the
state  $s_\beta$ assumed at the previous time $t_{n-1}$:  
\begin{align}
P(t_n, s=s_\alpha \mid t_{n-1},s = s_\beta; t_{n-2},s= s_\gamma; \dots )=\\ = P(t_n, s=s_\alpha \mid t =t_{n-1},s =s_\beta).
\end{align}
Then if the process is homogeneous in time,   it is completely characterized by the conditional probabilities 
$\hat{T}_{\alpha,\beta}=P(t_n, s=s_\alpha \mid t_{n-1, }s =s_\beta)$,.  The elements of such transition matrix cannot be completely  arbitrary: $0\le\hat{T}_{\alpha,\beta}\le1$  with $\sum_{\alpha}\hat{T}_{\alpha,\beta}=1$  because of  the normalization. 
\begin{figure}
	\centering
	\includegraphics[scale=0.3]{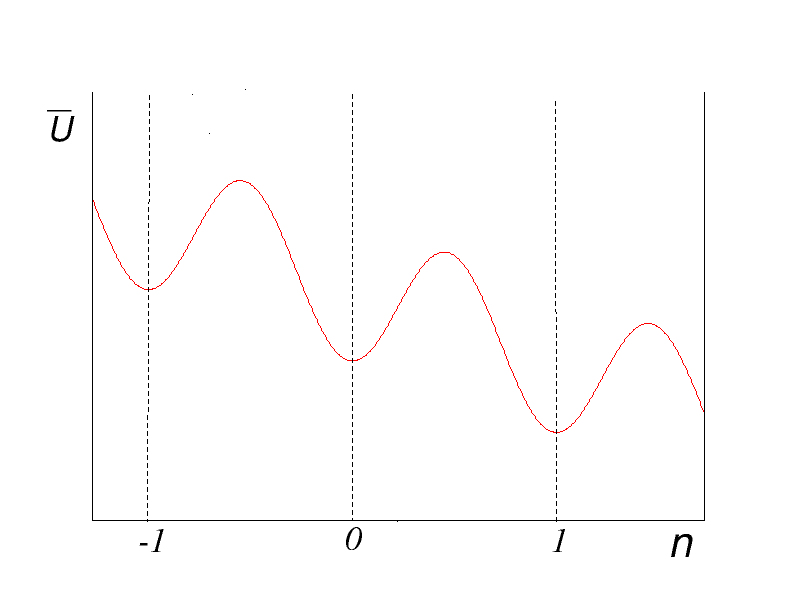}
	\caption{Schematic of the, one minimum, potential landscape  for the probe particle.}
	\label{potential}
\end{figure}

To describe effectively the stationary system dynamics, at each time step we associate  the particle coordinate to the closest  minimum of $\bar{U}(x)=U(x) - Fx$.   For sake of simplicity let us 
consider the simplest case in Fig. \ref{potential}  where 3 minima are shown.
For time intervals $\tau$ small enough,  the probability for the particle to perform a jump with $\mid \Delta n \mid  > 1$ is negligible. After a sampling time $\tau$, we again associate the coordinate of the particle to the nearest minimum of $\bar{U}(x)$. Then, we set the state of the Markov chain $ s = 1, -1,0 $ depending on whether the particle has jumped to the next, previous or current minimum respectively. It should be noted that the sampling time $\tau$ may differ from the integration time step. However, the results are weakly dependent on $\tau$ (at least for values of $\tau$ neither too large nor too small). The transition matrix  can be empirically estimated  by the ratio between the number of transitions $ N_ {ij} $ from  $i$ to   $j$, with $i,j=0,\pm1$,  and the number of times $N_i$ that  $i$ occurred :
\begin{equation} 
\hat{T}_{i, j} = \frac{N_ {ij}}{N_i} \quad \text{with} \quad i, j = -1,0,1 
\end{equation}
Being  the system periodic,  the transition probabilities $\hat {T} _ {i, j}$ are independent from the initial state $i$. By writing them  as:
\begin{equation} 
\hat {T}_{i , 1} = p_ + \quad \hat {T}_{i , 0} = p_0 \quad\hat{T}_{i , -1} = p_ -
\end{equation}
the stochastic matrix reduces to:
\begin{equation} 
\hat {T} = 
\begin{pmatrix} 
p _- & p_0 & p_ +  \\
p _- & p_0 & p_ + \\
p_-  & p_0 & p_ + 
\end{pmatrix} .
\end{equation}
The stationary probability  distribution satisfies 
$$
P^{\left\lbrace s \right\rbrace }=P^{\left\lbrace s \right\rbrace  } \hat{T},
$$
and corresponds therefore to the row eigenvector associated with the maximum eigenvalue of  $\hat{T}$, $\lambda = 1$:
\begin{equation}
P^{\left\lbrace s \right\rbrace } = \begin{pmatrix} p_-, \quad
p _0, \quad
p_+ 
\end{pmatrix} ,
\end{equation}
whose components can be  empirical estimated as
\begin{equation}
P^{\left\lbrace s \right\rbrace }_i = \frac {N_i}{\sum_ {j}{N_j}}.
\end{equation}
Now that we have introduced the general concepts, let's see how this works in practice. 

Consider the PT model, eq. (\ref{lang})  with constant external force  $F$ and a  periodic potential $U(x)$:
\begin{equation}
\label{oneminpot}
U(x)=\frac{U_0 L}{2\pi}\sin \biggl(\frac{2\pi}{L} x \biggr).
\end{equation}
Notice that Eq.~(\ref{lang})  can be made adimensional by introducing the natural units of mass $m$, time $m/\gamma$ and length $L$, turning into
\begin{equation}
\ddot{x}= F -\partial_x U(x)-\dot{x} +\sqrt{2T} \xi(t),
\label{alang}
\end{equation}
(where all quantities, $x$, $F$, $U$  are  now to be intended as expressed in the natural units), showing that 
therefore there are  only two free parameters, e.g.  $F/U_0$ and $T/U_0$. For the sake of clarity  in the following  we shall maintain the explicit  dependence on $L$.

To evaluate the transition matrix $\hat{T}$, we  resort to an integration algorithm from  \cite{manpal}. 
At each time step the  position $x (t)$ of the particle is recorded,  and  
after a sampling time step $\tau$ the closest minimum point of the potential is  determined. The state $s=1,-1,0$ is  associated to the particle depending if  the particle is in the basin of attraction of the next, the same, or the previous  potential minimum (notice that the state describes  the  transition and not to the visited minimum).

For all the numerical computations the maximum potential amplitude is set  to $U_0=0.1$ and the sampling time $\tau=0.1$, with  maximum computational  time $t_{max}=10^5$ (the dependence  of the results on the  value  of $\tau$ is very weak). Empirical averages  are estimated over $N=10^3$  different realizations of the noise with identical initial conditions for particle position and velocity.  As an example, for  $F=0.08$ and $T = 0.01 $ we  obtain:
\begin{equation} 
\hat{T} =
\begin{pmatrix}   0.00009  & 0.99554 & 0.00437 \\
 0.00009  & 0.99543 & 0.00448  \\
 0.00007 & 0.99529 & 0.00464 
\end{pmatrix}, 
\end{equation}
with little  fluctuations in the columns as consequence of numerical round-off  errors in the numerical integration process.

\begin{figure}
	\centering
	\includegraphics[scale=0.35]{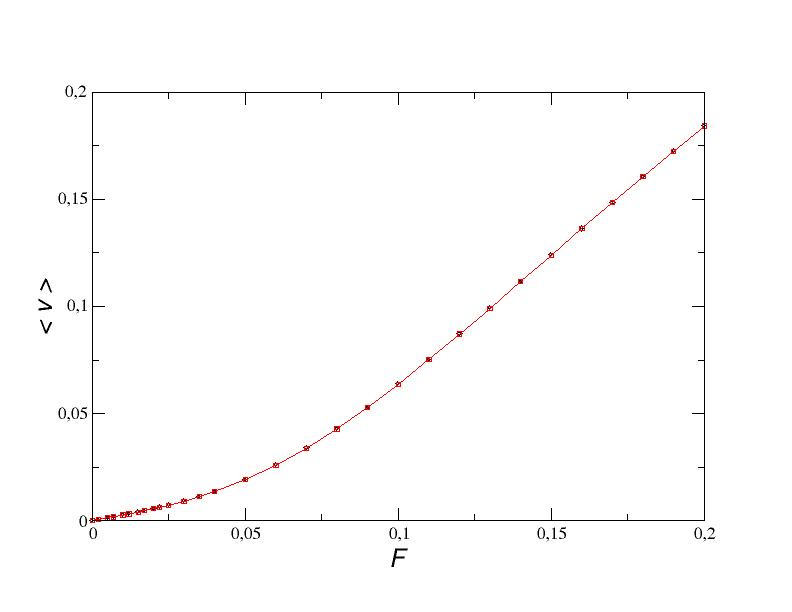}
	\caption{Comparison of  the average particle speed obtained from direct motion integration (circles, black on-line)  and from
		Markov chain description (squares, red  on-line) and  ) for a substrate potential with a single minimum per period, at $T=0.01$}
	\label{compspeed-onem}
\end{figure}

\section{Markov average velocity and friction}

As a test for the goodness of this  description we compare the average speed $\langle v \rangle$ obtained from the Markovian approximation 
\begin{equation}
\langle {v} \rangle = \frac{L}{\tau}(p_+ v_1 + p_0 v_0 + p_-v_{-1}) =  \frac{L}{\tau}(p_+ - p_-) 
\end{equation}
with that obtained  by direct integration of the motion equation, averaging on both time and  different realizations  of the process: 
\begin{equation}\bar{v} = \frac{1}{t_{max}} \sum_{i = 0}^{t_{max}} \langle v (t_i)\rangle,
\label{vtraiettoria}
\end{equation}
where $\langle \rangle$ stays here for  the average over the different realizations. 
Figure    \ref{compspeed-onem} shows that there is a very good agreement between the values obtained using the two methods  for different  values of the force $F$ at the temperature $ T = 0.01$   (with  relative errors  always smaller than $10^{-3}$).   
\begin{figure}
	\centering
	\includegraphics[scale=0.6]{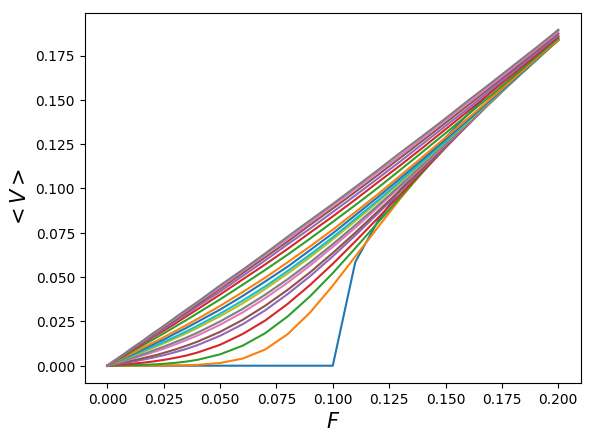}
	\caption{Average velocity as function of the external driving force in a one minimum potential for different $T$. The lowest curve (red on-line)  corresponds to $T=0$. The other curves, from bottom to top,  represent  increasing temperatures in  the range [0.003,0.05] corresponding  to those of the points in Fig. \ref{mT-onem}.}
	\label{v-onem}
\end{figure}

We then adopt the Markovian parametrization to investigate the behavior of $\langle v \rangle$ with $F$ at different temperatures. Results are shown in Fig.  \ref{v-onem}  As expected the curve for $T=0$ displays no motion as far as $F < U_0$ and is  singular in $F=U_0$. For $T > 0$ curves are smooth and display increasing values of  the average velocity with $T$. For increasing temperature   the average speed becomes essentially  proportional to $F$ since, as it will be shown below,  friction tends to vanish. At the same time thermal fluctuations become negligible for increasing $F$.  
\begin{figure} 
	\centering
			\includegraphics[scale=0.6]{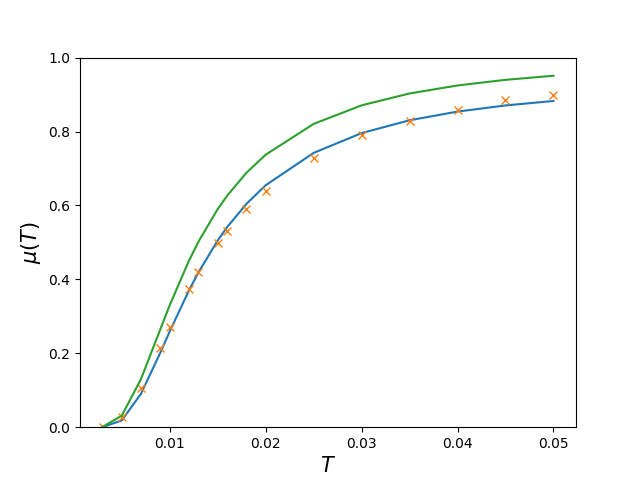}
	\caption{Mobility $\mu$ as  function of the bath temperature $T$ for a potential with one minimum as resulting from the Markov  chain (crosses). The upper curve (green on-line)  is obtained from the approximated expression derived  in \cite{Risken} in the high friction limit. The other curve (orange on-line) is best-fit to the points by  a generalization of the same expression (see text). }
	\label{mT-onem}
\end{figure}

For $F \ll U_0$,  on very general bases one expects for any $T$:
\begin{equation}
\label{mob}
\langle v \rangle =\mu \cdot  F,
\end{equation}
where $\mu=\mu(T)$ is the particle mobility.  Its values can be  obtained from the data of Fig. \ref{v-onem} by considering the values of   $\langle v \rangle $ at small $F$.  The behavior resulting from considering values below   $F=0.025 \ll U_0$ is shown by the points  in Fig. \ref{mT-onem}.
In the figure it is also reported the behavior resulting from the approximated derived in \cite{Risken} 
in the high friction limit (upper curve, green on-line)
\begin{equation}
\label{bessel}
\mu=\frac{I_0^{-2}(U_0/T)}{\gamma },
\end{equation}
where $I_0$ is the modified Bessel function. 
The lower curve is a best fit to the points obtained by inserting in an arbitrary proportionality factor between the two sides of the above equation and by letting it and $L$ to change  suitably.

Taking thermal averages in eq. (\ref{alang}),  in the steady state one has 
$\langle \ddot {x} \rangle=0$ and $\langle \xi (t) \rangle=0$, and the average speed is:
\begin{equation}
\langle v \rangle = \langle \dot {x} \rangle  = \frac{F - \langle \partial_xU (x) \rangle} {\gamma}. 
\label {v} 
\end{equation} 
Averages  are performed on the  probability distribution, that in turn depends on the force $F$, that can be determined  by the associate Fokker-Planck equation.
On the other hand in a deterministic system subjected to an average dynamic friction force $F_ {d}$ and a viscous coefficient 
$\gamma $, the stationary average speed of the particle is:
\begin{equation} 
\langle v \rangle  = \frac{F-F_ {d}}{\gamma}.
\label {vclassica} 
\end{equation} 
Thus one identifies the average value of the derivative of the interaction potential in  (\ref{v}) with the average dynamic friction\footnote{It can be useful to notice that in the literature  it is often assumed that at stationarity  $F$ equals the dynamic friction, that therefore in that   case incorporates the viscous term.}, which in turn depends on $F$:
\begin{equation} 
F_d (F) = \langle \partial_x U (x)\rangle \label{friction}.
\end{equation} 

\begin{figure}
	\centering
	\includegraphics[scale=0.4]{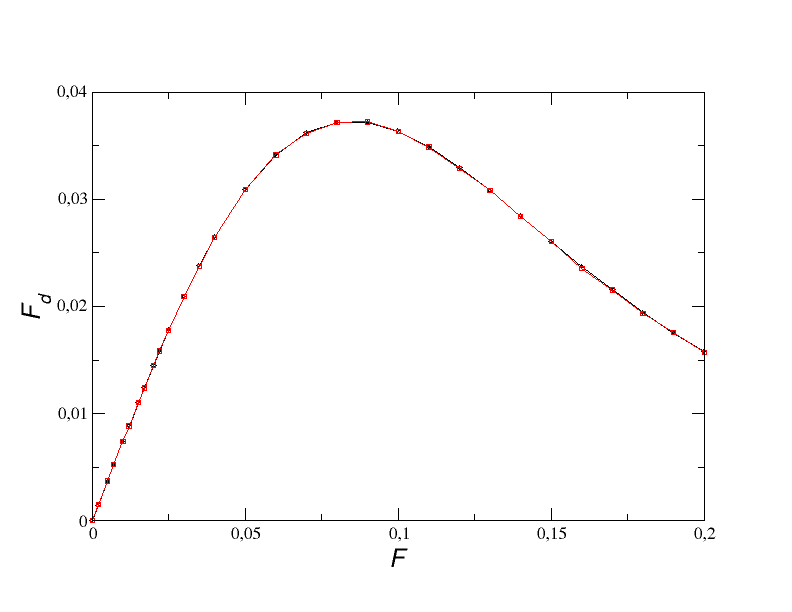}
	\caption{Comparison of the average  friction force acting on the particle, subjected to a single minimum periodic potential, as obtained from direct motion integration (circles, black on-line) and  from the  Markov chain, as difference between the average velocity and the driving  force  (squares, red on-line). $T=0.01$.}
	\label{compfric-onem}
\end{figure}

As a  further test for the Markovian description  we then compare the effective dynamic friction, eq. (\ref{friction}), resulting  from the Markov chain via eq. (\ref{v}) with that obtained by the numerical integration of motion. The two  are shown in Fig. \ref{compfric-onem}  displaying  even in this case a very good agreement  and relative  differences  smaller than $10^{-3}$. As expected $F_d$  displays a maximum in correspondence of the value of $F=U_0$, and decreases for larger  values, so that $\langle v\rangle \to F/\gamma$. 
\begin{figure}
	\centering
	\includegraphics[scale=0.6]{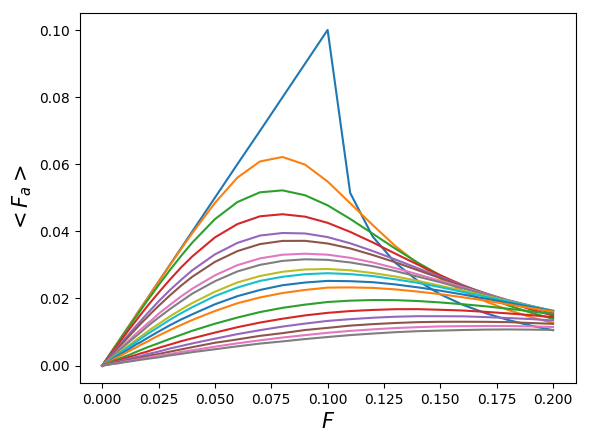}
	\caption{Average dynamic friction, in a single minimum potential, as function of the external driving force for different $T$. The upmost curve (red on-line) corresponds to $T=0$. The other curves represent temperatures  in the range [0.003,0.05],  decreasing from top to bottom, and correspond to the temperatures of the points in Fig. \ref{mT-onem}.}
	\label{Fd-onem}
\end{figure}

Figure  \ref{Fd-onem} displays $F_d$, computed via the stochastic matrix, as function of $F$  for different temperatures  ranging from $T=0$ to $T=0.05$.  For   $T=0$, corresponding  to the uppermost curve,  the average friction equals the driving force, with a discontinuous derivative, indicating static equilibrium for any value  $F \le U_0$, and corresponding therefore to  a static friction.
For $T \neq0$  the particle can  jump from a local potential well to another one for any value of $F$ and $U_0$. If  $F=0$ the probability of moving forward to the next  minimum  is  equal to that of moving backward, and the average speed is zero. 
When $F \neq 0$,  its direction is favored and the average speed is different from zero, implying  $\langle\partial_x U (x)\rangle< F$ even if $F \ll U_0$. 
As expected, the general effect of thermal fluctuation is to decrease friction, whose values tend to become similar 
for $F  \gg U_0$.   It is however interesting to notice that the  minimum friction at large $F$ is obtained for  $T=0$ .

\section{Entropy production}

One  advantage of  the description in terms of a  Markov chain is that it allows to define and compute in a simple way  the entropy production of the process along system's trajectories. The only required condition is that if
$\hat {T} _ {i, j}> 0$, then also $\hat {T}_ {j, i}> 0$.  Within this frame the  probability of observing  a trajectory  of length $n$ starting  from the state  $s_0$ and ending  in $s_n$, $ \{s_0;  s_1;  \dots; s_n \}$,  can be expressed as 
\begin{equation}
\mathit{P}(s_0;\dots ; s_n)=\hat{T}_{n,n-1},\dots\hat{T}_{2,1}\hat{T}_{1,0}\pi_{0}
\end{equation}
and similarly, the probability of the reverse trajectory is:
\begin{equation}
\mathit{P}(s_n; \dots; s_0)=
\hat{T}_{0,1}\dots\hat{T}_{n-2,n-1}\hat{T}_{n-1,n}\pi_{n}
\end{equation}
where $ \pi_0 $ and  $ \pi_n$ are the  invariant probabilities  of $s_0$ and $s_n$.
The entropy production along the  path  can then be computed as \cite{Maes1999,leb}:
\begin{equation}\begin{split}\Sigma_n&=\ln{\biggl(\frac{\mathit{P}(s_0; \dots; s_n)}{\mathit{P}(s_0; \dots ; s_n)}\biggr)}-\ln{\biggl(\frac{\pi_{0}}{\pi_{n}}\biggr)}=\\
&=\sum_{k=1}^{n}\ln{\biggl(\frac{\hat{T}_{\alpha_k,\alpha_{k-1}}}{\hat{T}_{\alpha_{k-1},\alpha_k}}\biggr)}\end{split}.
\end{equation}
\begin{figure}
	\centering
	\includegraphics[scale=0.6]{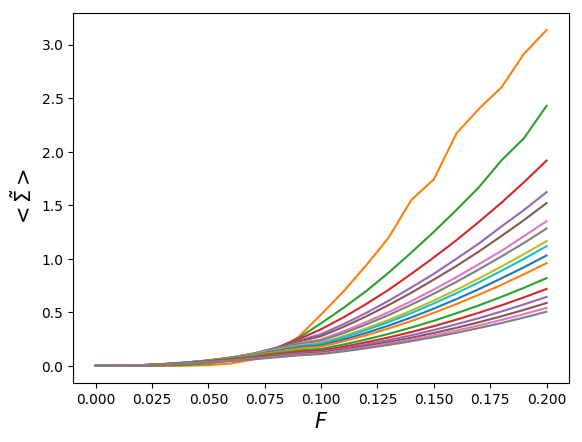}
	\caption{Average rate of entropy production $<\tilde{\Sigma}>$, for the case of periodic potential with a single minimum, as function of the driving force $F$ at  temperatures corresponding to those of the points of Fig.  \ref{cTunminimo}, with  increasing values when going from  top to  bottom.}
	\label{entropiaunminimo}
\end{figure}

To calculate it  we observe that  the position of the particle at the time $t_n$  is related  to the states of  the Markov chain through:
\begin{equation}
x_n=x_{n-1}+\delta_{n}
\end{equation}
where  $\delta_n$ assumes the values: $-L,0,L$ 
with probabilities  $p_+$, $p_-$ e $p_0$, respectively.
If  at time $n = 0$ the position of the particle $x_0$ is  in the state $\alpha_0$,  the probability of observing a given trajectory $ \{x_0; x_1; \dots; x_n \} $. 
depends therefore  exclusively on the probability  of the initial state and on number of times, $n _+ $, $ n_-$ and $n_0$,  that  $\delta_k$ takes the value $L$, $-L$ and $0$:
\begin{equation}
\mathit{P}(x_0;\dots;x_n)=\pi_{0}p_+^{n_+}p_-^{n_-}p_0^{n_0} 
\end{equation}
The probability for the inverse trajectory is obtained by exchanging the number of forward and backward transitions  $n_+$ and $n_-$:  
\begin{equation}
\mathit{P}(x_n;\dots;x_0)=\pi_{n}p_+^{n_-}p_-^{n_+}p_0^{n_0}, 
\end{equation}
and the entropy production results in 
\begin{equation}
\Sigma=\ln{\biggl(\frac{p_+^{n_+}p_-^{n_-}p_0^{n_0}}{p_+^{n_-}p_-^{n_+}p_0^{n_0}}\biggr)}=(n_+-n_-)\ln{\biggl(\frac{p_+}{p_-}\biggr)} .
\end{equation}
In the limit $n \gg 1$, the number of forward jumps $n_+$ and backward jumps $n_-$  tends to their expected  values,  $np _+$ and $np _-$, and the  entropy production rate $<\tilde{\Sigma}>=\frac{\Sigma}{n\tau}$ assumes the average  value
\begin{equation}
\langle \tilde{\Sigma}\rangle=
\frac{1}{\tau}(p_+-p_-)\ln{\biggl(\frac{p_+}{p_-}\biggr)} \label{sigmaunabuca}.
\end{equation}

\begin{figure} 
	\centering
		\includegraphics[scale=0.6]{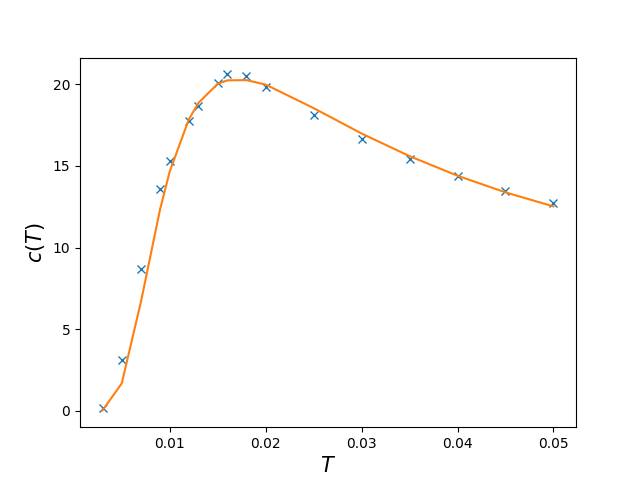}
	\caption{Proportionality constant $c(T)$  between the entropy production rate and $F^2$ at low $F$ for one-minimum potential.}
	\label{cTunminimo}
\end{figure}

The  resulting behavior as function of $F$ is shown  in Fig.  (\ref{entropiaunminimo}) for various temperatures between  $T=0.003 $ (the upper curve, red on-line) and  $T= 0.05$ (the lower curve, dark pink  on-line).  As thermal fluctuations become less relevant,  the entropy production increases  because of the decrease of $p_-$.The behavior for  $F \ll U_0 $ can be understood by observing that in this case probabilities $p_+$ and $p_-$ do not differ too much one  each other and can be written as:
\begin{align}
&p_+=p+\epsilon_1\\
&p_-=p-\epsilon_2
\end{align}
with  $\epsilon_1, \epsilon_2 \ll 1$.
By expanding  the logarithm in (\ref{sigmaunabuca}),  and considering that  $\langle  v\rangle =
\frac{L}{\tau}(\epsilon_1+\epsilon_2)$, one then obtains:
\begin{equation}
\langle\tilde{\Sigma}\rangle\simeq \frac{(\epsilon_1+\epsilon_2)^2}{p\tau}=\frac{<v>^2	\tau}{pL^2} \label{sigmaFpiccolo},
\end{equation}
or
\begin{equation}
\langle\tilde{\Sigma}\rangle\simeq c(T)F^2,
\end{equation}	
with 
\begin{equation}
c(T)=\frac{\tau}{L^2p} \, \mu^2(T).
\label{cTviamu}
\end{equation} 
The  behavior of $c(T)$  is  shown in Fig.$\ref{cTunminimo}$ and  displays a relative maximum. 
The continuous (orange) curve is obtained by a fit with the expression eq. (\ref{cTviamu}).

Finally let us observe that  the entropy production is proportional  to  the current:
\begin{equation}
J_n=\sum_{k=1}^{n}{\delta_k} ,
\end{equation}
an interesting quantity  for systems far from the equilibrium. 

Let us stress that it is possible to adopt a Markov description in the analysis of experimental data even in absence of a detailed model. In fact, it is enough to identify the relevant variables which describe the systems. Furthermore, it allows to define and compute in a relatively simple way the entropy production of the system avoiding the complications that arise in the continuous description due to the non-invertibility of the noise matrix. 

\section {Corrugated potential}

\begin{figure} 
	\centering
	\includegraphics[scale=0.3]{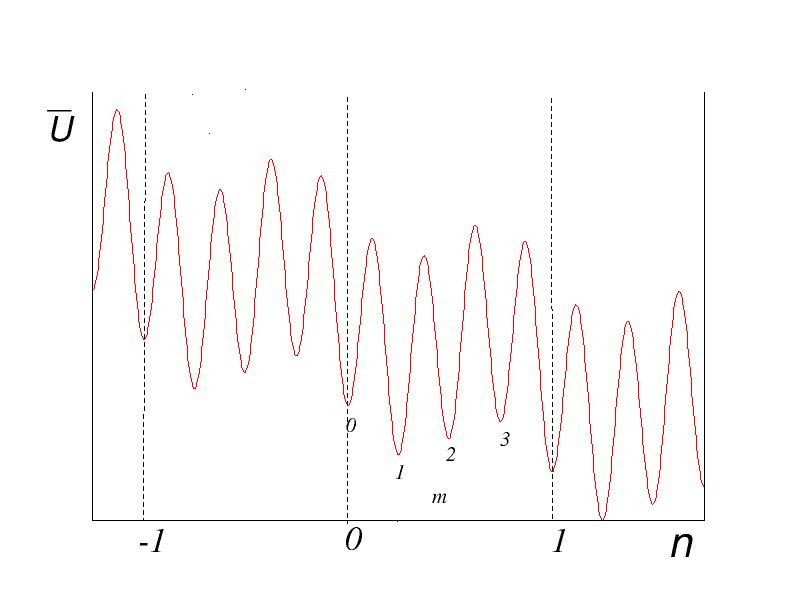}
	\caption{Schematic of the corrugated potential landscape for the probe particle. The small numbers indicate the states $s$ of the Markov chain.}
	\label{fourminpotential}
\end{figure}

The above analyses can be easily repeated also for  the case of a potential  displaying several minima within each period. One way to obtain such potential is to add to $U(x)$  another periodic term  with amplitude $\Delta_0$, and period, $l$:
\begin{equation}
U(x)=\frac{U_0L}{2\pi}\sin\biggl(\frac{2\pi}{L}x\biggr)+\frac{\Delta_0l}{2\pi}\sin\biggl(\frac{2\pi}{l}x\biggr), \label{potnew}
\end{equation}
with  $L= ml$. The number of relative minima  depends on the choice of $\Delta_0$ and ranges between $0$ and $m$ . As an instance, we choose $l=L/4$, and $\Delta_0=3U_0$,
so   $U(x)$  displays four  minima within one  period $L$ and there are four equilibrium points determined by the condition
\begin{equation} 
F-U_0\cos\biggl(\frac{2\pi}{L}x\biggr)-\Delta_0\cos\biggl(\frac{8\pi}{L}x\biggr)=0. \label{stabilita}
\end{equation}
In this case, the states of the Markov chain $s=0,1,2,3$ correspond to the basins of attraction of the relative minima of the potential (see Fig. \ref{fourminpotential}). 

\begin{figure} 
	\centering
	\includegraphics[scale=0.6]{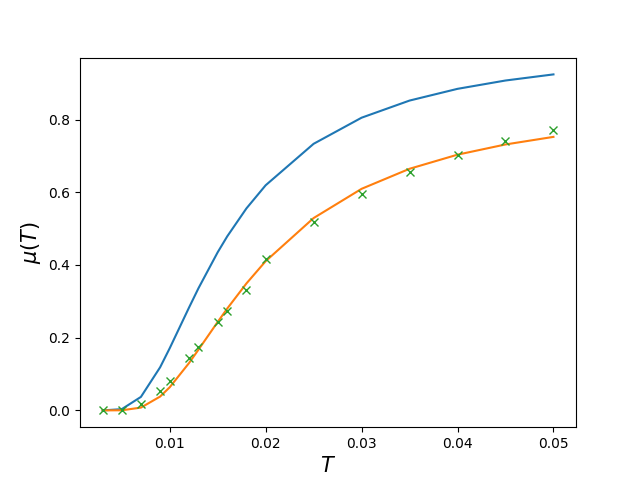}
	\caption{Mobility $\mu$ as  function of the bath temperature $T$ for a potential with four  minima (crosses).  The upper curve  (blue on-line) is obtained from the approximate expression derived in \cite{Risken}. The lower curve  (orange on-line) is given by a best-fit  of a generalization of this expression  (see text) to the points. }
	\label{mT-piuminimi}
\end{figure}

Although in this case also transitions between different, not necessarily neighboring,  secondary minima are  allowed,  
the observed dependence of the average friction $F_d$  and speed on  the external force $F$ and  temperature $T$
results qualitatively  very similar to the case of a single minimum, as shown in Fig. \ref{mT-piuminimi}.  The figure also shows  (upper curve, green on-line)
the behavior as resulting from the high friction limit expression \cite{Risken}:
\begin{equation}
\mu = \left( \frac{2\pi}{\gamma\int_0^{2\pi} e^{U/T} dx}\right)^2.
\end{equation}
The lower curve (orange on-line) is   obtained by  fitting  of the points with  the same  expression but adjusting  $L$ and an arbitrary prefactor.
More interesting appears the behavior of the entropy. In this case,  simple manipulations  lead to the following expression for   the average rate of entropy  production over a trajectory of length $t=n\tau $:
\begin{equation}
\langle \tilde{\Sigma}\rangle =\frac{1}{n \tau}\sum_{i=0}^3{(\pi_i\hat{T}_{i,i+1}-\pi_{i+1}\hat{T}_{i+1,i})\log{\biggl(\frac{\hat{T}_{i,i+1}}{\hat{T}_{i+1,i}}\biggr)}} \label{sigmaunitapiubuche}
\end{equation}
where $\pi_i$ is the stationary  probability of observing the particle close to  the $i$th  secondary minimum  ($i=0,1,2,3$). 

\begin{figure}
	\centering
	\includegraphics[scale=0.6]{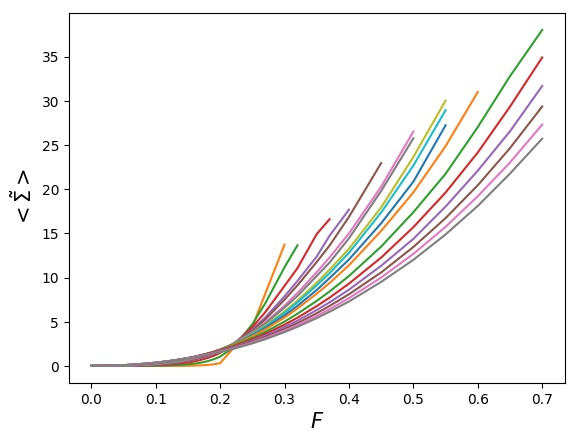}
	\caption{Average entropy production rate $<\tilde{\Sigma}>$ as function of the driving $F$   for a  four-minima potential. The different curves correspond  to the temperatures of the points in Fig.  \ref{cTpiuminimi},  with   increasing  values from the shortest  to the longest. For large $F$ inverse trajectories are forbidden  at low $T$.}
	\label{entropiapiuminimi}
\end{figure}

\begin{figure} 
	\centering
		\includegraphics[scale=0.6]{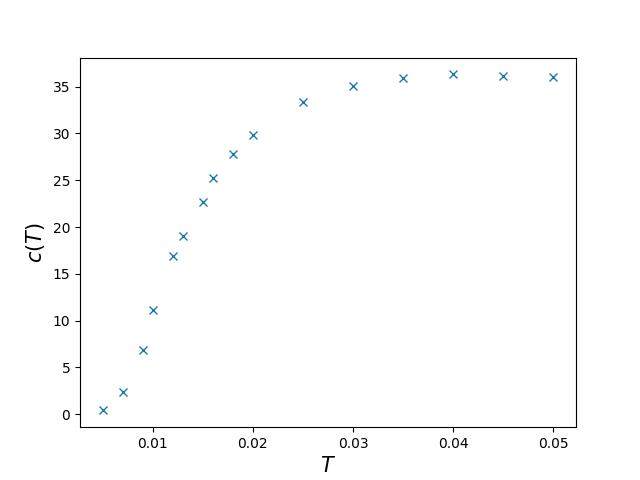}
	\caption{Proportionality constant $c(T)$  between the entropy production rate and $F^2$  at low $F$ for a four-minima potential.}
	\label{cTpiuminimi}
\end{figure}

The resulting dependence on $F$ at different $T$ is shown in Fig. \ref{entropiapiuminimi}. 
The first thing  that  can be noted is that the entropy production cannot be calculated for each value of the force $F$ since at low temperature  thermal fluctuations are not  enough strong to allow the particle to jump backward for large   $F$.  On the other hand, in analogy to  the case of a single minimum,   $\langle \tilde{\Sigma} \rangle $  is proportional to $F^2$ for small $F$.
The resulting proportionality coefficient   $c(T)$ is  shown  in Fig.  (\ref{cTpiuminimi}). 
At variance with the one-minimum potential the maximum is much less pronounced.

\section{Summary}

In this paper we have investigated the stationary state of the inertial Prandtl-Tomlinson model in a thermal bath describing it by  a suitable Markov chain. The description in terms of a Markov chain is rather flexible and can be used even in presence of lack of an accurate model: it suffices to know the proper variables. 

We have considered in particular the less  investigated instance of a constant driving force,  considering  the periodic potential in both the case of a single minimum and  of  many minima per period. 
At variance with other approaches \cite{Furlong2009,Braun2010,Perez2010,Muser2011,Dong2012,Xu2019}  we have not estimated the transition rates from Arrhenius or  Kramers  expressions   but by  direct numerical simulations of the  system's dynamics, as described by the corresponding Langevin equation.  Under the assumption that at each sampling time step  the probe particle could  jump  at most the distance of one period of the potential,   we have computed  friction and  velocity  from the Markov chain   for a wide range temperatures $T$ and driving forces $F$, finding excellent agreement with the values obtained by direct simulations. 

With  this approach we have then been able to investigate also the  behavior of the entropy production. To our knowledge this has been studied  only  recently by Torche  and coworkers \cite{Torche2019} for the case of a compliant driving,
starting from a master equation with Arrhenius transition rates.
We have considered the periodic potential in both  the case of  one single minimum and of many minima per period. We have  derived an explicit expression for the case of a single minimum  at low temperature,  finding  that in this limit the entropy production is proportional to the square of the external force, and that  the proportionality coefficient  displays  a non monotonic behavior as function of the temperature.

\bibliography{PT}

\begin{thebibliography}{10}
\expandafter\ifx\csname url\endcsname\relax
  \def\url#1{\texttt{#1}}\fi
\expandafter\ifx\csname urlprefix\endcsname\relax\def\urlprefix{URL }\fi
\expandafter\ifx\csname href\endcsname\relax
  \def\href#1#2{#2} \def\path#1{#1}\fi

\bibitem{Dong2011}
Y.~Dong, A.~Vadakkepatt, A.~Martini, {Analytical models for atomic friction},
  Tribology Letters 44~(3) (2011) 367--386.
\newblock \href {https://doi.org/10.1007/s11249-011-9850-2}
  {\path{doi:10.1007/s11249-011-9850-2}}.

\bibitem{Vanossi2013}
A.~Vanossi, N.~Manini, M.~Urbakh, S.~Zapperi, E.~Tosatti, {Colloquium: Modeling
  friction: From nanoscale to mesoscale}, Reviews of Modern Physics 85~(2)
  (2013) 529--552.
\newblock \href {http://arxiv.org/abs/1112.3234} {\path{arXiv:1112.3234}},
  \href {https://doi.org/10.1103/RevModPhys.85.529}
  {\path{doi:10.1103/RevModPhys.85.529}}.

\bibitem{Popov2012}
V.~L. Popov, J.~A. Gray, {Prandtl-Tomlinson model: History and applications in
  friction, plasticity, and nanotechnologies}, ZAMM Zeitschrift fur Angewandte
  Mathematik und Mechanik 92~(9) (2012) 683--708.
\newblock \href {https://doi.org/10.1002/zamm.201200097}
  {\path{doi:10.1002/zamm.201200097}}.

\bibitem{Prandtl1928}
L.~Prandtl,
  \href{https://www.reibungsphysik.tu-berlin.de/fileadmin/fg204/Publikationen/Prandtl_1928_English.pdf}{{Ein
  Gedankenmodell zur kinetischen Theorie der festen K\"orper}}, Zeitschrift fr
  Angewandte 8 (1928) 85, {E}nglish translation: A Conceptual Model to the
  Kinetic Theory of Solid Bodies, Translated from German original by V. L.
  Popov and J. Gray, Berlin University of Technology.
\newline\urlprefix\url{https://www.reibungsphysik.tu-berlin.de/fileadmin/fg204/Publikationen/Prandtl_1928_English.pdf}

\bibitem{Tomlinson1929}
G.~Tomlinson, A molecular theory of friction, Philos. Mag. 7 (1929) 905–939.

\bibitem{spikes2015}
H.~Spikes, W.~Tysoe, {On the Commonality between Theoretical Models for Fluid
  and Solid Friction, Wear and Tribochemistry}, Tribology Letters 59~(1) (2015)
  1--14.
\newblock \href {https://doi.org/10.1007/s11249-015-0544-z}
  {\path{doi:10.1007/s11249-015-0544-z}}.

\bibitem{Fusco2005}
C.~Fusco, A.~Fasolino, {Velocity dependence of atomic-scale friction: A
  comparative study of the one- and two-dimensional Tomlinson model}, Physical
  Review B - Condensed Matter and Materials Physics 71~(4) (2005) 1--9.
\newblock \href {http://arxiv.org/abs/0502496} {\path{arXiv:0502496}}, \href
  {https://doi.org/10.1103/PhysRevB.71.045413}
  {\path{doi:10.1103/PhysRevB.71.045413}}.

\bibitem{Nakamura2005}
J.~Nakamura, S.~Wakunami, A.~Natori, {Double-slip mechanism in atomic-scale
  friction: Tomlinson model at finite temperatures}, Physical Review B -
  Condensed Matter and Materials Physics 72~(23) (2005) 1--5.
\newblock \href {https://doi.org/10.1103/PhysRevB.72.235415}
  {\path{doi:10.1103/PhysRevB.72.235415}}.

\bibitem{Furlong2009}
O.~J. Furlong, S.~J. Manzi, V.~D. Pereyra, V.~Bustos, W.~T. Tysoe, {Kinetic
  Monte Carlo theory of sliding friction}, Physical Review B 80~(15) (2009)
  153408.
\newblock \href {https://doi.org/10.1103/PhysRevB.80.153408}
  {\path{doi:10.1103/PhysRevB.80.153408}}.

\bibitem{Dong2012}
Y.~Dong, D.~Perez, H.~Gao, A.~Martini, {Thermal activation in atomic friction:
  Revisiting the theoretical analysis}, Journal of Physics Condensed Matter
  24~(26) (2012).
\newblock \href {http://arxiv.org/abs/NIHMS150003} {\path{arXiv:NIHMS150003}},
  \href {https://doi.org/10.1088/0953-8984/24/26/265001}
  {\path{doi:10.1088/0953-8984/24/26/265001}}.

\bibitem{Wang2015}
Z.~J. Wang, T.~B. Ma, Y.~Z. Hu, L.~Xu, H.~Wang, {Energy dissipation of
  atomic-scale friction based on one-dimensional Prandtl-Tomlinson model},
  Friction 3~(2) (2015) 170--182.
\newblock \href {https://doi.org/10.1007/s40544-015-0086-2}
  {\path{doi:10.1007/s40544-015-0086-2}}.

\bibitem{Furlong2015}
O.~J. Furlong, S.~J. Manzi, A.~Martini, W.~T. Tysoe, {Influence of Potential
  Shape on Constant-Force Atomic-Scale Sliding Friction Models}, Tribology
  Letters 60~(2) (2015) 1--9.
\newblock \href {https://doi.org/10.1007/s11249-015-0599-x}
  {\path{doi:10.1007/s11249-015-0599-x}}.

\bibitem{Xu2019}
R.~G. Xu, Y.~Xiang, Q.~Rao, Y.~Leng, On the asymptotic expressions of critical
  energy barrier in prandtl-tomlinson model, International Journal of Smart and
  Nano Materials 10~(2) (2019) 107--115.

\bibitem{bowman2013introduction}
G.~R. Bowman, V.~S. Pande, F.~No{\'e}, An introduction to Markov state models
  and their application to long timescale molecular simulation, Vol. 797,
  Springer Science \& Business Media, 2013.

\bibitem{Norris}
J.~Norris, Markov Chains, Cambridge, 1997.

\bibitem{manpal}
R.~Mannella, V.~V.~Palleschi, Fast and precise algorithm for computer
  simulations of stochastic differential equations, Phys. Rev. A 40 (1989)
  3381.

\bibitem{Risken}
H.~Risken, The Fokker-Planck Equation, Springer-Verlag, 1989.

\bibitem{Maes1999}
C.~Maes, The fluctuation theorem as a gibbs property, J. stat. Phys. 95 (1999)
  67–392.

\bibitem{leb}
J.~L. Lebowitz, H.~Spohn, A {G}allavotti-{C}ohen-type simmetry in the large
  deviation functional for stochastic dynamics, Journal of Statistical Physics
  95 (1999).

\bibitem{Braun2010}
O.~M. Braun, M.~Peyrard, {Master equation approach to friction at the
  mesoscale}, Physical Review E - Statistical, Nonlinear, and Soft Matter
  Physics 82~(3) (2010).

\bibitem{Perez2010}
D.~Perez, Y.~Dong, A.~Martini, A.~F. Voter, {Rate theory description of atomic
  stick-slip friction}, Physical Review B - Condensed Matter and Materials
  Physics 81~(24) (2010) 1--6.
\newblock \href {https://doi.org/10.1103/PhysRevB.81.245415}
  {\path{doi:10.1103/PhysRevB.81.245415}}.

\bibitem{Muser2011}
M.~M\"user, {Velocity dependence of kinetic friction in the Prandtl-Tomlinson
  model}, Physical Review B - Condensed Matter and Materials Physics 84~(12)
  (2011).
\newblock \href {https://doi.org/10.1103/PhysRevB.84.125419}
  {\path{doi:10.1103/PhysRevB.84.125419}}.

\bibitem{Torche2019}
P.~C. Torche, T.~Polcar, O.~Hovorka, Thermodynamic aspects of nanoscale
  friction, Physical Review B 100~(12) (2019) 125431.

\end{thebibliography}

\end{document}